\documentclass[iop,numberedappendix,twocolappendix]{emulateapj}
\usepackage{mathptmx}
\usepackage{fancyvrb}
\usepackage{epstopdf}

\usepackage{amsmath}
\usepackage{amssymb}

\usepackage{graphicx}
\usepackage{float}
\usepackage{natbib}
\usepackage{footnote}
\usepackage{bbold}

\DefineVerbatimEnvironment{code}{Verbatim}{fontsize=\small}
\DefineVerbatimEnvironment{example}{Verbatim}{fontsize=\small}

\def\gtorder{\mathrel{\raise.3ex\hbox{$>$}\mkern-14mu
             \lower0.6ex\hbox{$\sim$}}}
\def\ltorder{\mathrel{\raise.3ex\hbox{$<$}\mkern-14mu
             \lower0.6ex\hbox{$\sim$}}}

\newcommand{\argmax}{\arg\!\max}

\submitted{Submitted to ApJ}

\shorttitle{Optimal source detection and photometry}

\shortauthors{Zackay \& Ofek}

\begin{document}




\title{How to coadd images? I. Optimal source detection and photometry using ensembles of images}
\author{Barak Zackay\altaffilmark{1} and
Eran O. Ofek\altaffilmark{1}
}
\altaffiltext{1}{Department of Particle Physics and Astrophysics, Weizmann Institute
  of Science, Rehovot, Israel}

\begin{abstract}

Stacks of digital astronomical images are combined in order to increase image depth.
The variable seeing conditions, sky background and transparency
of ground-based observations make the coaddition
process non-trivial.
We present image coaddition methods optimized for 
source detection and flux measurement, that maximize the signal-to-noise ratio ($S/N$).
We show that for these purposes the best way to combine images
is to apply a matched filter to each image using
its own point spread function (PSF) and only then to sum the images with the appropriate weights.
Methods that either match filter after coaddition, or perform PSF homogenization prior to coaddition will result in loss of sensitivity.
We argue that our method provides an increase of between
a few and 25 percent in the survey speed of deep ground-based imaging surveys compared with weighted coaddition techniques. We demonstrate this claim using simulated data as well as data from the Palomar Transient Factory data release 2.
We present a variant of this coaddition method which is optimal for PSF or aperture photometry.
We also provide an analytic formula for calculating the $S/N$
for PSF photometry on single or multiple observations.
In the next paper in this series we present a
method for image coaddition in the limit of background-dominated
noise which is optimal for any statistical test or measurement on the constant-in-time image
(e.g., source detection, shape or flux measurement or star-galaxy separation), making the original data redundant.
We provide an implementation of this algorithm in MATLAB.

\end{abstract}


\section{Introduction}
\label{sec:Introduction}
Coaddition of observations is one of the most basic  
operations on astronomical images.
The main objectives of coaddition are to
increase the sensitivity (depth) of observations,
to remove signal from high energy particle hits on the detector
(sometimes called cosmic rays),
and to decrease the point spread function (PSF) width
(e.g., deconvolution from wavefront sensing, \citealp{DWFS}; lucky imaging, \citealp{Law06}).
One question that is common for all these tasks is what
is the best method to coadd the images. This of course requires
a definition of what "best" means.

Here, we assume that all observations are registered and re-sampled to the same grid (e.g., using {\tt Swarp}; \citealp{Bertin2002}), and focus on the combination operation of the images.
There are several common complications for image coaddition.
Ground-based images
are often taken under variable seeing, background and transparency conditions.
Even some space-based observations may suffer from such problems.
For example, X-ray images commonly have non-uniform PSF accross the field of view.
These complications make the coaddition operation less trivial
than simple (weighted) scalar addition.

Coaddition is playing, and will play, a major role in ongoing and future surveys
(e.g., SDSS, \citealp{SDSS};
Pan-STARRS, \citealp{PANSTARRS};
PTF, \citealp{PTF}, 
DES, \citealp{DES};
LSST, \citealp{LSST}).
There are many approaches for image coaddition
after the registration step.
One pre-step common to many methods is to reject
some of the images with the worst seeing, background or transparency.
There is no sensible prescription to which images one should reject, and it is straightforward to show that this approach leads to a loss of sensitivity compared with the maximum possible. One extreme example for this process is lucky imaging \citep{Law06}, where one often discards up to 99\% of the data frames.
The next coaddition steps involves either weighted summation
of the images, or applying partial filtering (or deconvolution) to each image
in order to homogenize the PSFs of all the images followed
by coaddition.

For example, \cite{Annis2014} recommended coadding the SDSS Stripe 82 images
by weighted summation of the images, where the weights are of the form:
\begin{align}
w_{j} = \frac{F_{j}^{k}}{s_{j}^{m}V_{j}^{n}}\,,
\label{eq:Annis}
\end{align}
and $k=1$, $m=2$, and $n=1$.
Here $w_{j}$ is the weight applied to the image $i$,
$F_j$ is proportional to the product of the telescope effective area,
detector sensitivity and atmospheric transparency (i.e., flux-based photometric zero point),
$s$ is the width of the PSF,
and $V$ is the variance of all the background noise sources
(e.g., background, readout noise).
On the other hand \cite{Jiang2014} adopted a similar
formula but with $k=1$, $m=1$, $n=1$.
Another example for such a weighted summation was given by \cite{Fischer94} who solved a numerical optimization scheme for the optimal weights in which the images should be added with. 

Other approaches are also used.
For example, some authors perform PSF homogenization on all
the images prior to coaddition. This is done by convolving the frames with some kernel in order to bring all the observations to have the same PSF. For example, \cite{Darnell2009} and \cite{Sevilla2011} define a median seeing PSF, and then for each image find the convolution kernel that will transform 
the image (via convolution) to a median-seeing image,
applying for all images, and then sum the images.
This convolution kernel can be found using techniques outlined
in \cite{Phillips95}, \cite{AL98} and \cite{Bramich}.
However, for images with seeing which is worse than the target seeing, the PSF homogenization operation becomes a deconvolution operation, which amplifies the noise in high spatial frequencies and creates long range correlations in the noise.
Another issue of the PSF homogenization approach
is the artificial insertion of correlations between neighboring pixels. As a result, any measurement on the data becomes more involved. For example, applying a naive matched filter with the coadd-image PSF will result in an additional loss of information.
%

In a series of several papers we will tackle the problem
of optimal image coaddition.
In this paper (paper~I), we will lay down the statistical formalism and assumptions we use for the coaddition problem. We construct a coaddition method that achieves the maximum possible signal-to-noise ratio ($S/N$) for detection of non-blended faint point-like sources in the coadded image under the assumption of constant variance Gaussian noise\footnote{The constant noise requirement can be relaxed as one can solve the problem locally to a region in
which the noise variance is roughly constant.}.
In addition, we present a coaddition method that is optimal for photometric measurements of both faint and bright objects (i.e., even for sources in which the source noise is non-negligible).
These coaddition products are the analogues of
linear matched filter for ensemble of images. We note that this coaddition
method, optimized for source detection, was used by \cite{MasciFowler2008}
to reduce the {\it WISE} satellite data.
However, in order to recover the sharpness of the original images they applied de-convolution after the coaddition process.

It is noteworthy that linear matched filter images are smeared and their pixels are correlated. Therefore, general hypothesis testing and statistical measurements on them requires knowledge and consideration of the spatial covariance matrix.
This makes them non-attractive for visualization and non intuitive for performing measurements other than source detection and flux measurement (e.g., resolving binary stars or measuring galaxy shapes).

In Paper~II \citep{Coad2} we build
on the results of this paper to construct a coaddition method
for the special (but common) case of background-dominated noise.
Equipped with the background-dominated noise assumption
we find a simple transformation that
removes the pixel correlation induced by the matched-filtering step,
and makes the additive noise in all pixels uncorrelated and with equal variance.
Most importantly,
this technique provides a sufficient statistic\footnote{A statistic is sufficient with respect to a model and its associated parameter if no other statistic that can be calculated from the same sample provides any additional information as to the value of the models parameter.} for {\it any} statistical measurement
or decision on the data (e.g., source detection, flux measurement,
shape measurement).
Moreover, this method conserves the information of all the spatial frequencies,
and provides a PSF which is typically sharper than
the PSF of the best image in the stack we coadd.
In addition, this image is practically indistinguishable from a regular astronomical observation, and any image analysis code can be applied to it unchanged.
Combining all these properties together, it allows a major reduction in the amount of data products users will need in order to perform any further processing on data (e.g., galaxy shape measurements for weak lensing).
Last, this coaddition algorithm is fast and its implementation is trivial.

In future papers (Zackay et al., in prep.) we apply the algorithm from Paper~II for the case of rapid imaging through the turbulent atmosphere and where the PSF is either known (through observing a reference star or using a wavefront sensor) or unknown.
We show that this method potentially provides better results than the existing approaches for the coaddition of speckle images (e.g., speckle interferometry, \citealp{SpeckleInterferometry}; lucky imaging, \citealp{Law06}, and its suggested improvements, e.g., \citealp{Garrel12}). 

The structure of this paper is as follows:
In \S \ref{sec:StatisticalBackground}, we review the statistical formalism of image coaddition and refer to some basic lemmas regarding how to combine random variables.
In \S \ref{sec:DetectionFaint}, we discuss the optimal coaddition image for faint source detection.
In \S \ref{sec:SourcePhotometry}
we describe how to obtain the optimal coaddition image for photometric measurements, not restricted to faint sources.
In \S \ref{sec:TheoreticalSensitivity}, we give a simple to use formula to calculate the expected signal-to-noise ratio for source detection and PSF photometry for both single images and ensembles of images.
In \S \ref{sec:WeightedMedian}, we suggest possible variants of this method, that albeit sub-optimal for source detection and photometry, are resilient to particle hits, and generalize the concept of median for images.
In \S\ref{sec:tests}, we demonstrate our detection algorithm
on simulated and real images and show it surpasses the
sensitivity of current popular methods.
In \S\ref{code} we briefly describe the code we use.
Finally, we summarize the important formulae and conclude in \S \ref{sec:Disc}.

\section{Statistical formalism and background}

\label{sec:StatisticalBackground}

Denote the $j$'th measured image, in a series of background-subtracted observations of the same position by $M_j$. Further, denote by $P_j$ the point spread function (PSF) of the $j$'th observation.

The statistical model for the $j$'th image is given by:
\begin{align}\label{eq:M}M_j = (F_jT)\otimes P_j + \epsilon_j,\end{align}
where $T$ denotes the true (background subtracted) sky image, $F_j$ is a scalar, representing the transparency (same as in \S \ref{sec:Introduction}), $\otimes$ represents convolution, and the noise term $\epsilon_j$ has variance \begin{align}V[\epsilon_j] = B_j + (F_jT)\otimes P_j\,,\end{align} where $B_j$ is the variance of all the position independent noise -- the sum of the sky background variance, the read noise variance and the dark current variance.
For simplicity, we assume that the gain is 1 and that the counts are measured in electrons. 

In the following subsections we derive several well known
important results of signal detection using hypothesis testing.
The optimal detection of an attenuated signal in the presence of varying noise is reviewed in \S\ref{subsec:ExampleGaussians}.
In \S\ref{subsec:WeightedAddition} we derive a general rule for weighted addition of random variables, while in \S\ref{subsec:MatchedFilter} we derive the well known linear matched filter solution for source detection.

\subsection{Detection and measurement of an attenuated signal in the presence of varying noise}\label{subsec:ExampleGaussians}

In this section we review the simple problem
of how to detect or measure a signal, with a known template, in data given by an ensemble of observations. We strictly assume that the signal in all observations is attenuated linearly with known attenuation coefficients, and that the noise in all the observations is Gaussian.
Albeit simple, the solution to this problem is the key towards constructing an optimal algorithm for image coaddition.

Let $X_j$ be a set of independent Gaussian variables with variance 
\begin{align}
V[X_j] \equiv \sigma_j^2\,.
\end{align}
Given the null hypothesis, which states that there is no signal, and is denoted by $\mathcal{H}_0$, we assume: \begin{align}E[X_j|\mathcal{H}_0] = 0\,,\end{align}
where $E[X|\mathcal{H}]$ denotes the expectancy of the variable $X$ given that the hypothesis  $\mathcal{H}$ is true.
Given the alternative hypothesis, that states there is a signal, and is denoted by $\mathcal{H}_1(T)$, we assume: 
\begin{align}E[X_j|\mathcal{H}_1(T)] = \mu_jT\,.\end{align} Here $\mu_j$ are known scaling factors of each observation.
The log-likelihood ratio test statistic (which is proven to be the most powerful test at any given size, and therefore optimal. See \citealp{NeymanPearsonLemma}), is simply given by:

\begin{align}
\mathcal{L} = \sum_j{\frac{X_j^2}{2\sigma_j^2}} - \sum_j{\frac{(X_j-\mu_jT)^2}{2\sigma_j^2}}\,.
\end{align}
For detection purposes, i.e., to reject the null hypothesis, the term $\frac{\mu_j^2T^2}{2\sigma_j^2}$ can be ignored, as it does not depend on the data. However, this term is still relevant for flux measurements.
Simplifying and absorbing the factor of 2 into the likelihood, we get:
\begin{align}
 \mathcal{L}= \sum_j{\frac{\mu_jT}{\sigma_j^2}X_j}\,.
\end{align}
The log-likelihood ratio is linear in $T$, hence we can identify a sufficient statistic
\begin{align}\label{eq:S}
S = \sum_j{\frac{\mu_j}{\sigma_j^2}X_j}\,,
\end{align}
with a distribution that is independent of $T$, and therefore can be used to reject the null hypothesis against the alternative hypotheses $\mathcal{H}_1(T)$ for all values of $T$ simultaneously. Therefore, the optimal strategy for detection of a statistically significant signal, is to calculate $S$, and check if it is above or below a certain threshold $S_\gamma$. The value of $S_{\gamma}$ is fixed by the desired false alarm probability $\gamma$.
To evaluate this threshold, we need to calculate the expectancy and variance of $S$ given $\mathcal{H}_0$:
\begin{align}
E[S|\mathcal{H}_0] = 0\,,
\end{align}
Given the independence of $X_j$ and the fact that $V[X_j|\mathcal{H}_0]=\sigma_j^2$, the variance of $S$ given $\mathcal{H}_0$ is:  
\begin{align}\label{eq:I} V[S|\mathcal{H}_0] = \sum_j{\frac{\mu_j^2}{\sigma_j^2}}\equiv I\,.\end{align}
And the threshold is:\footnote{This threshold was calculated under the convenient approximation that $\gamma =\int\limits_{x}^{\infty}{\frac{e^{-x^2/2}}{\sqrt{2\pi}}}\approx \frac{1}{x\sqrt{\pi}}e^{-x^2/2}\approx e^{-x^2/2 - 2}$, which is a good approximation for $4<x<8$, which is the reasonable significance range for practical surveys.} \begin{align}S_\gamma \simeq \sqrt{-2I\ln(\gamma) - 4}\,.\end{align}

Under the same definitions, if we wish to get the most accurate estimator of $T$, we can use the fact that the likelihood of the data can be calculated for all $T$ using $S$ (Eq. \ref{eq:S}) and $I$ (Eq. \ref{eq:I}).
This is because we can express the log likelihood of $M$ given $T$ by:
\begin{align}
\mathcal{L}(M|T) = \sum_j\frac{(X_j-\mu_jT)^2}{2\sigma_j^2} =g(x) - ST +\frac{IT^2}{2}\,,
\end{align}
where $g(X) = \sum_j{\frac{X_j^2}{2\sigma_j^2}}$ does not depend on $T$ and therefore cannot influence any measurement.
No matter if the Bayesian or frequentist approaches are used, the likelihood of the data as a function of $T$ is sufficient for any further measurement or detection process. Therefore, $S$ and $I$ are sufficient for any measurement or statistical decision on the set of statistics $\{X_j\}$.

\subsection{Weighted addition of random variables}\label{subsec:WeightedAddition}

The statistics $S$ and $I$ we have derived in Equations \ref{eq:S} and \ref{eq:I} naturally arise in a much more general setup, which is known as weighted addition of random variables.
Assume we have an arbitrary set of independent random variables $X_j$ and two hypotheses $\mathcal{H}_0,\mathcal{H}_1$ that satisfy \begin{align}\label{Cond:EqualVariance}V[X_j|\mathcal{H}_0] = V[X_j|\mathcal{H}_1] \equiv V(X_j).\end{align} 
We can define the signal to noise ratio ($S/N$) of a statistic $S$ with regard to the decision between $\mathcal{H}_0$ and $\mathcal{H}_1$ by: \begin{align}S/N[S] = \frac{E(S|\mathcal{H}_0) - E(S|\mathcal{H}_1)}{\sqrt{V(S)}}\,.\end{align}

Next, we can ask: what is the statistic, linear in $X_{j}$, that maximizes this $S/N$?
The solution to this problem is the following weighted mean:
\begin{align}\label{eq:S_proper_add}S = \sum_j{\frac{E(X_j|\mathcal{H}_0) - E(X_j|\mathcal{H}_1)}{V(X_j)}X_j}\,,\end{align}
which coincides with the previously defined $S$ (Eq. \ref{eq:S}).
Moreover, the maximal $S/N$ is $\sqrt{I}$.
The detailed solution of this proplem is given in Appendices \ref{AppendixProperAddition} and \ref{AppendixMaxSNR}.


Weighted addition extends also to the case of adding estimators. If we are given a sequence of independent measurements 
\begin{align}
\theta_j = \mu_jT + \epsilon_j\,,
\end{align} in which we assume only that $V[\epsilon_j] = \sigma_j^2$, The same set of statistics $S$ and $I$ arise, and the maximum S/N estimator for $T$ is :
\begin{align}\label{eq:ProperFluxEstimate}
\tilde{T} = \frac{\sum_j{\frac{\mu_j\theta_j}{\sigma_j^2}}}{\sum_j{\frac{\mu_j^2}{\sigma_j^2}}} = \frac{S}{I}\,.
\end{align}
This simple and rigorous fact is derived in Appendix \ref{AppendixProperAdditionEstimators}.

The fact that this same solution repeats under many different general contexts and assumptions, allows us to derive optimal statistics to real-world astronomical problems. The first of them is the well known matched filter for source detection in single astronomical images.

\subsection{The matched filter solution for source detection}\label{subsec:MatchedFilter}
Now that we are equipped with the simple yet robust tools for adding random variables, we can derive the standard matched filter solution for point source detection in a single image.
To allow rigorous statistical analysis of source detection, we assume (throughout the paper) that the source we are trying to detect is well separated from other sources, which might interfere with its detection.
Let $M$ be an image such that:
\begin{align}
M = T\otimes P + \epsilon\,,
\end{align}
where $T$ is the true, background-subtracted sky image, $P$ is the image PSF, and $\epsilon$ is an additive white Gaussian noise term.

We would like to detect, or measure, a single point source
at position ($x_{0}$,$y_{0}$).
Each pixel coordinates ($x$,$y$) in the image is
an independent\footnote{In practice the pixels maybe slightly correlated due to charge repulsion and charge diffusion in a CCD.} information source, where the PSF scales the amount of flux each pixel would get. That is:
\begin{align}
\mu(x,y) = P(x-x_0,y-y_0)\,.
\end{align}
Using equation \ref{eq:S}, we can write the optimal statistic for source detection at position $(x_0,y_0)$ for a single image:
\begin{align}
S(x_0,y_0) &= \sum_{x,y}{\frac{P(x-x_0,y-y_0)M(x,y)}{\sigma_j^2}} \\&= \left(\frac{\overleftarrow{P}\otimes M}{\sigma_j^2}\right)(x_0,y_0)\,,
\end{align}
where $\overleftarrow{P}(x,y) = P(-x,-y)$.
Thus, we obtained the well known optimal statistic for the detection of a single source in an image with additive white Gaussian noise. This solution is to convolve the image with it's reversed PSF (i.e., cross correlation of the image with it's PSF). This statistic is called the linear matched filter image.

Due to its optimality, robustness and ease of implementation and interpretation,
the linear matched filter is the method of choice for many
source detection tools
(e.g., {\tt SExtractor}; \citealp{Bertin96};
see a review in \citealp{SourceDetectionReview}).

\section{Image coaddition for source detection}
\label{sec:DetectionFaint}

Now that we have shown that using the tools outlined in \S\ref{sec:StatisticalBackground} we get the optimal statistic for source detection in single images, we can apply these methods for source detection in an ensemble of images.
Before doing that it is worth while to develop an intuition
on how the solution should look like.
Expanding the linear matched filter solution of one image to
multiple images we expect the solution to be somewhat like
a three dimensional linear matched filter summed over the image index axis, to get a statistic with the dimensions of an image.
Therefore, we expect that the solution will be simply
matched filtering each image with its own PSF followed by weighted summation.

For source detection, we assume that the noise is approximately Gaussian
and that the variance in each pixel, excluding
the variance induced by the source itself,
is independent of pixel position.
Since our null hypothesis is that there is no source,
the noise induced by the source should be ignored.
However, noise induced by neighboring sources can~not be ignored.
Therefore, our assumption that the variance is constant
(at least locally) in the image is true only if the image is not crowded\footnote{For detecting a source in the vicinity of other sources, our estimator is obviously biased by the nearby source, which is not accounted for in the statistical model presented in this work. Correctly resolving and measuring blended sources requires a more elaborate technique in general. In paper II in this series, we show a coaddition process that is optimal for all hypothesis testing cases, including resolving blended sources.}
(i.e., confusion noise can be neglected).
Using these assumptions, we denote the variance of the $j$'th
image by:
\begin{align}
\sigma_{j}^{2} \equiv B_j =  V[\epsilon_{j}]\,.
\label{eq:Vb}
\end{align}
Here $B_j$ is the background (or more formally the spatially invariant variance component) of the $M_j$ image.
We can now apply the machinery for coaddition of random variables developed in \S\ref{sec:StatisticalBackground} to the problem of image coaddition.

Lets assume that we would like to detect a point source
at position $(0,0)$.  In this case the model for such a point source
is $T=T_{0}\delta_{(0,0)}$, where $T_0$ is the point source's flux and $\delta_{(0,0)}$ is a two dimensional array with $1$ at position $(0,0)$ and zero otherwise. Inserting this to the model of $M_j$ we get:
\begin{align}M_j = (F_jT_0\delta_{(0,0)})\otimes P_j + \epsilon_j = F_jT_0P_j  + \epsilon_j\,.\end{align}

Using the result from Appendix \ref{AppendixProperAddition}, we can combine all the pixels from all the images as a list of independent statistics. Therefore the optimal detection statistic for a source at position $(0,0)$ is:
\begin{align}
S = \sum_{j,x}{\frac{F_jP_{x,j}}{\sigma_j^2}M_j(x)}\,.
\end{align}
For simplicity $x$ represents the $(x,y)$ coordinate.
Using the fact that searching a source at position $p \equiv (p_1,p_2)$ on a set of images, is exactly as searching for a source at position $(0,0)$ on the input images shifted by $-p$ we get:

\begin{align} \label{eq:S_all}
S(p) = \sum_{j,x}{\frac{F_jP_{x,j}}{\sigma_j^2}M_j(x-p)}.
\end{align}
This can be written in vectorial notation as:
\begin{align} \label{eq:S_conv}
S = \sum_j{ \left(\frac{F_j\overleftarrow{P_j}}{\sigma_j^2}\right)\otimes M_j}\,,
\end{align}
where $\overleftarrow{P_j}(x) = {P_j}(-x)$.
We note that this could be easily computed (using the fact that $\sigma_j^2$ is a scalar) in Fourier space using the fast Fourier transform:
\begin{align}\label{eq:S_hat}
\widehat{S} = \sum_j{\frac{F_j\overline{\widehat{P_j}}}{\sigma_j^2}\widehat{M}_j}\,,
\end{align}
where the $\widehat{\;\;}$ sign represents Fourier transform,
and the $\overline{\raisebox{1.5mm}{} \quad}$ sign denotes the complex conjugate operation.
We note that by putting the bar sign over the hat sign
we mean that the complex conjugate operation
follows the Fourier transform operation.

It is worth while to note that $S$, the coaddition of filtered images is a log likelihood ratio image, where $S(p_1,p_2)$ is the optimal test statistic for the existence of a source at $(p_1,p_2)$.
In order to find sources using $S$, we therefore have to search for local maxima.
To decide if the source candidate is statistically significant, it is more convenient to normalize $S$ by it's standard deviation
\begin{align}\label{eq:S_sig}
S_{\sigma} = \frac{S}{\sigma_S} =  \frac{\sum_j{\left(\frac{F_j\overleftarrow{P_j}}{\sigma_j^2}\right)\otimes M_j}}{\sqrt{\sum_{j,x}{\frac{\overleftarrow{P_{j,x}}^2F_j^2}{\sigma_j^2}}}}\,.
\end{align}
The value of each local maximum divided by the standard deviation of the noise (Eq. \ref{eq:S_sig}) is the source significance in units of standard deviations.
To get $S$ in units of flux (best estimate only in the context of background-noise dominated source), we need to normalize differently (see Eq.~\ref{eq:ProperFluxEstimate}), to get:

\begin{align}
S_{\rm flux} = \frac{\sum_j{\left(\frac{F_j\overleftarrow{P_j}}{\sigma_j^2}\right)\otimes M_j}}{\sum_{j,x}{\frac{F_j^2\overleftarrow{P_{(j,x)}}^2}{\sigma_j^2}}}\,.
\end{align}

Analyzing our solution we can see that if we have only one image then
the well known linear-matched filter is recovered.
However, when we have more than one image, the optimal detection statistic is equivalent to applying a matched filter to each image separately, weighting the images by $\frac{F_j}{\sigma_j^2}$, and accumulating. This is different than the other solutions found in the literature, where you first coadd and then match filter, or perform partial filtering for homogenization, then coadd, and then match filter.
The method we developed is optimal by construction,
hence it will have better performance relative to other coaddition schemes.
We note that with this method at hand, any additional image,
no matter how poor the conditions are, will increase the sensitivity.
Therefore, the rejection of images becomes a damaging operation.
Because of the robustness of our optimality arguments, whenever the assumptions that we have made on the noise are true we expect this method to surpass the performance of all popular methods found in the literature.

An important question is by how much our method
improves source detection relative to other methods.
The answer to this question depends on the details, like the distribution of observing conditions (e.g., seeing) and the exact method we would like to compare with.
We can quantify the gain of using our method, by answering a simple question: how much survey speed is being lost
when a non-optimal filter is used in the case of a single image?
A single matched filter cannot be adequate for multiple
images with different PSFs. No matter what weights are later applied in the combination of the different matched filtered images (all with the same filter), the $(S/N)^2$ of the other methods cannot surpass the sum of the $(S/N)^2$ in their individual pixels. 

Given a single image with a Gaussian PSF with width $s$,
the best filter is a Gaussian with width $s$.
In Appendix~\ref{Ap:InfoPSFlost} we calculate the $S/N$ loss factor,
when one uses a Gaussian filter with a width $ks$.
We find that the sensitivity ratio of these two cases
is:
\begin{align}\frac{\left(S/N\right)_{\rm incorrect\;\;width}}{\left(S/N\right)_{\rm optimal}}=\left(\frac{2k}{k^2+1}\right).\end{align}
This means that the reduction in survey speed (or the detection information) is:
\begin{align}\label{eq:SurveySpeedIncrease} \frac{\left(S/N\right)_{\rm incorrect\;\;width}^2}{\left(S/N\right)_{\rm optimal}^2} = \left(\frac{2k}{k^2+1}\right)^2.\end{align}

\begin{figure}[h]
\centering
\includegraphics[width = 85mm]{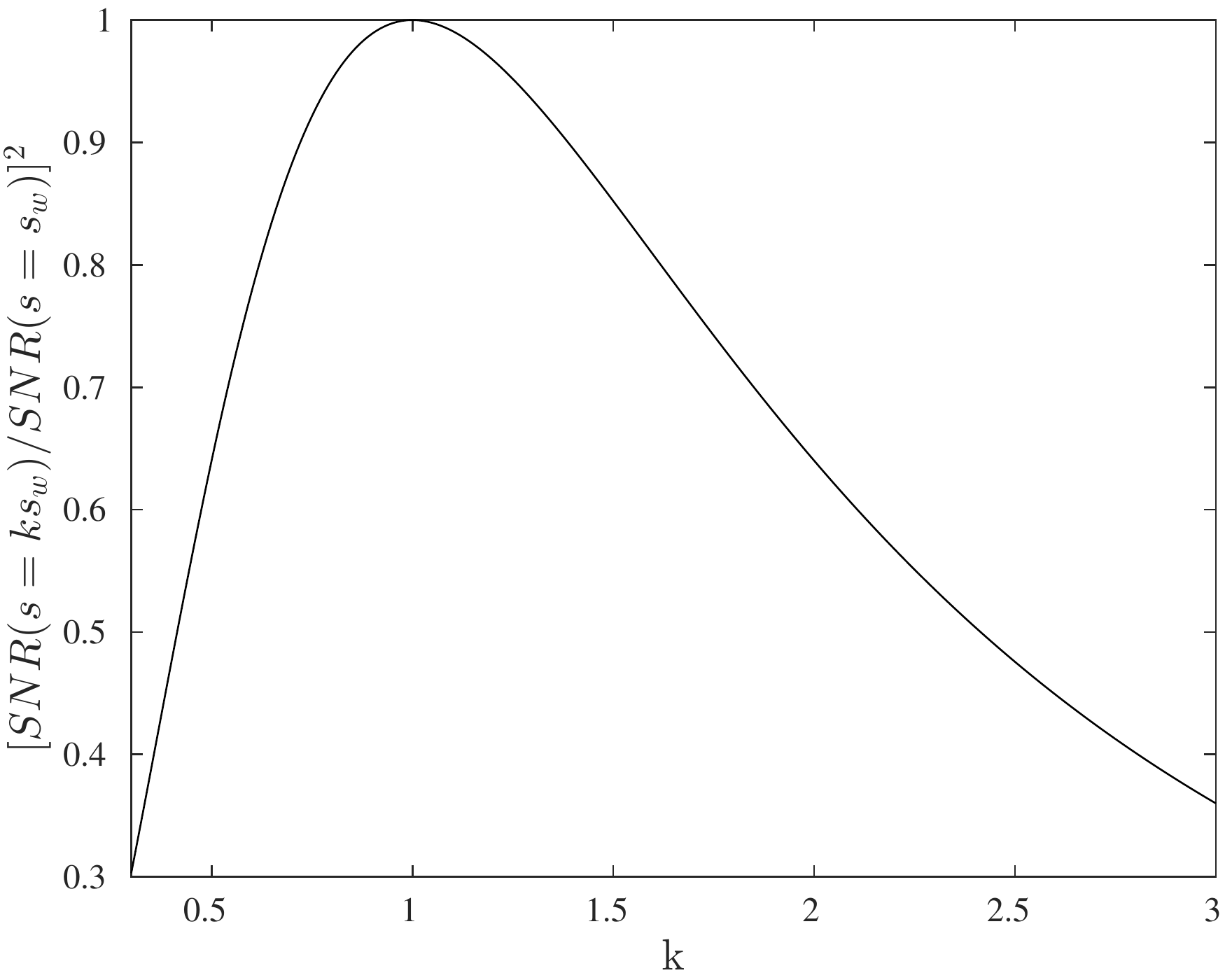}
\caption{Survey speed loss due to the use of a matched filter which width is $k$ times that of the optimal filter (Eq.~\ref{eq:SurveySpeedIncrease}).\label{fig:SurveySpeed}}
\end{figure}

Figure~\ref{fig:SurveySpeed} shows this survey speed reduction as a function of k.
As expected, this function has a maximum of unity at $k=1$.
Using a width which is a factor of $1.2$ ($2$) different than
the actual PSF width will result in survey speed loss of about 6\% (40\%).
Therefore, given the typical distribution of seeing conditions,
one should expect to obtain an improvement of between a few percents and about 25 percents in survey speed using our method, relative to e.g., the \cite{Annis2014} weighted summation.

\section{Optimal photometric measurements from a set of observations}
\label{sec:SourcePhotometry}

Another important problem in the context of astronomical image processing is measuring the flux of sources.
The two leading methods for performing flux measurements are aperture photometry and PSF photometry.
Both methods are well defined on single images.
When the PSF is exactly known, the method of PSF photometry is the most exact since it applies an appropriate weight to each pixel before summation. Aperture photometry is a good approximation that is often used when either the star is bright, the PSF is unknown, or the measurement is done for sources that are not point like.
In this section, we present the correct way to extend photometric measurements to ensembles of observations.
In \S\ref{subsec:PSFphot} we discuss the extension of PSF photometry to ensembles of images and in \S\ref{subsec:AperPhot} we cover aperture photometry.
We note that in the case of spatially homogeneous noise variance (i.e., faint source) and known PSF, it is both easier and more accurate, to use the coaddition method presented in paper~II, and then perform aperture/PSF photometry. 

\subsection{Extending PSF photometry to ensembles of images}\label{subsec:PSFphot}

We define the optimal photometric measurement as a statistic $S$ such that the expectancy of the statistic is the true flux of the source and its variance is minimal. That is,
$E[S] = T\,,$ and $V[S]$ is minimal.
This is equivalent to finding a statistic that maximizes the signal-to-noise ratio, and normalizing it to units of flux.
Because in the case of photometry we need to take into account the source noise, in this section we will work under the assumption that:
\begin{align}
M_j = F_jT\otimes P_j + \epsilon_j, 
\end{align} where the distribution of $\epsilon_j$ is Gaussian with variance \begin{align}V[\epsilon_j] = B_j + F_jT\otimes P_j\,,\end{align}which does not distribute uniformly across the image, because bright stars have larger variance components.
We note that we still assume that the measured source is well separated from other sources.

Following the same procedure we did for source detection, we get the maximum $S/N$ for the flux measurement.
When measuring a point source at position $(0,0)$, with true flux $T_0$ we can write:
\begin{align}M_j = (F_jT_0\delta_{(0,0)})\otimes P_j + \epsilon_j = F_j P_j + \epsilon_j\,.\end{align}
Again, we assume every pixel is statistically independent from the others, and therefore we can apply the previously developed machinery (Eq. \ref{eq:ProperFluxEstimate}) pixel by pixel, and get the maximal $S/N$ estimator:
\begin{align}\label{eq:T_PSF_phot}
\tilde{T}_{\rm PSF\;photometry} = \frac{ \sum_{j,x}{\frac{F_jP_{x,j}}{B_j + F_jP_{x,j}T_0}M_j(x)}}{\sum_{j,x}{\frac{F_j^2P_{x,j}^2}{B_j + F_jP_{x,j}T_0}}}\,.
\end{align}

However, in order to compute this statistic, we need to know the true flux $T_0$ that we are looking for. This problem could be avoided by either:
\begin{itemize} 
\item Solving numerically for the true $T_0$ in iterations:
Guess $T_0 = 0$, compute $\tilde{T}_{\rm PSF\;photometry}$, and then update $T_0=\tilde{T}_{\rm PSF\;photometry}$, and compute $\tilde{T}_{\rm PSF\;photometry}$ again with the resulting flux from the first iteration. Since the first estimate of $T_0$ is an excellent first guess, the latter is practically converged, and more iterations are unnecessary. This solution is source specific, and therefore should be implemented only if photometry of a single source is needed.
\item Scanning the $T_0$ space logarithmically - Using a slightly wrong $T_0$ does not lead to significant errors in the measurement.
Since the guess of $T_0$ influences only weights that are independent of the data itself, it does not introduce a bias.
The standard deviation of a statistic that is calculated with a slightly wrong $T_0$ is generally very close to the optimal value. For example, using Gaussian PSFs with width $1<s<10$ pixels, $1<B<1000$ and $1<T<10^{6}$, we find that using $T_{0}$ which is wrong by a factor of 2, leads to reduction of at most 1\% in survey speed.
This solution requires the co-addition process to produce several images ($\sim 8$, as the relevant range is from the background level up to the maximal dynamic range value. These numbers are typically only up to a factor of $2^8$ from one another).
\item Aim for a specific $T_0$. If one knows that the targets have a specific magnitude range, it is possible to use the most relevant $T_0$.
\end{itemize}
We want to stress, that the whole discussion on $T_0$ is only relevant for flux estimation of bright sources.
For the detection of sources, we are not required to know $T_{0}$ apriori, as demonstrated in \S \ref{sec:DetectionFaint}.

The calculation of this measurement statistic can be done fast, by noticing that once $T_0$ is assumed, the calculation of the measurement statistic on all image positions can be done by convolution using the fast Fourier transform (FFT)\footnote{Calculation of convolution directly (without FFT) could also be very fast if the convolution kernel is small.}.

To gain intuitive interpretation of the PSF photometry measurement statistic, it is useful to inspect the behavior of the statistic on one image.
In the expression for $\tilde{T}_{\rm PSF\;photometry}$, we can identify the two general approximations astronomers often make in the extreme limits of the relation between the brightness of the target and the background. On the one hand, if a target is brighter than the background on many pixels, we can see that the optimal measurement statistic we obtained is similar to aperture photometry with the optimal aperture, as in the center the weights of the pixels converge to 1. On the other hand, we see that if a target is much weaker than the background variance, the statistic converges to a matched filter on the image.
When extending this to multiple images we can see that in the statistic $\tilde{T}_{\rm PSF\;photometry}$, the combination of different images is rather non-trivial. 

\subsection{Extending aperture photometry for ensembles of images}\label{subsec:AperPhot}
Even when the PSF is unknown, or when the object shape for which we perform flux measurements is poorly known (e.g., a galaxy), it is still important to combine the flux measurements from different images correctly. This is because observations in the ensemble will generally have different noise properties. For each image, depending on the seeing conditions, source flux and its PSF, one can choose the best aperture radius $r_j$ that maximizes the $S/N$ in the image (see below). 

We can calculate from each image the aperture photometry measurement \begin{align}\label{eq:theta}\theta_j = M_j\otimes C_{r_j}\,,\end{align}where $C_{r_j}$ is given by:
\begin{align}
 &C_{r_j} =1 \quad{\rm if} \quad |r| < r_j \\ \nonumber
 &C_{r_j} =0 \quad{\rm if} \quad |r| \ge r_j.
\end{align}
Substituting Equation~\ref{eq:M} into Equation~\ref{eq:theta}:
\begin{align}\theta_j = F_jT\otimes P_j \otimes C_{r_j} + \epsilon_j\otimes C_{r_j}\,.\end{align}
Denote the source intrinsic light distribution by $Q$.
$Q$ satisfies:
\begin{align}\sum_x{Q(x)} = 1\,, \end{align} and $T = T_0Q$.
Next, we define 
\begin{align}
\mu_j = F_jQ\otimes P_j\otimes C_{r_j}\,,
\end{align} 
to be the fraction of the source flux within the measurement aperture.
Now, we can write:
\begin{align}
\theta_j = \mu_j T_0 + \epsilon_j\otimes C_{r_j} \,.
\end{align}
The variance of $\epsilon_j\otimes C_{r_j}$ is:
\begin{align}
V[\epsilon_j\otimes C_{r_j}] = \mu_jT_0 + \pi r_j^2\sigma_j^2\,.
\end{align}

Now that the problem is cast into the same form as solved in \S \ref{sec:StatisticalBackground}, we can use the maximum $S/N$ measurement statistic (Eq.\ref{eq:ProperFluxEstimate}) to find the best estimator for aperture photometry:
\begin{align}\label{eq:Theta_aper_phot}
\tilde{\theta}_{\rm aperture\; photometry} = \frac{\sum_j\frac{\mu_j}{\mu_jT_0 + \pi r_j^2\sigma_j^2}\theta_j}{\sum_j\frac{\mu_j^2}{\mu_jT_0 + \pi r_j^2\sigma_j^2}}\,.
\end{align}

Note that if the source is very strong, the term $\mu_jT_0$ might dominate the noise variance $V[\epsilon_j\otimes C_{r_j}]$.
In that case, one needs to have a reasonable guess of the source flux $T_0$ prior to the coaddition process (see \S\ref{subsec:PSFphot}).

Moreover, one needs to solve for the best aperture radius $r_j$ (that depends on $T_0$). This solution can be obtained separately for each image by maximizing the squared signal to noise of $\theta_j$:
\begin{align}
r_j = \argmax_r\left(\frac{\mu_j^2(r)}{\mu_jT_0 + \pi r^2\sigma_j^2}\right)\,,
\end{align}
where $\argmax_r(f(r))$ denotes the value of the parameter $r$ for which the function $f(r)$ obtains its maximum.
Maximizing this could be done if one has a model for $\mu(r)$, which could be deduced from the intrinsic size of the object measured and a rough PSF model.


\section{Estimating the limiting flux and the expected S/N of sources}
\label{sec:TheoreticalSensitivity}

For several reasons, including $S/N$ calculation
and optimizations, it is worth while to
have a formula for the $S/N$ of PSF photometry over
multiple images.
Albeit this formula is easy to derive, we are not aware of a simple analytic expression for the $S/N$ of PSF photometry in the literature for either single or multiple observations.

Following the previous section, we do not necessarily assume that the sources are weak compared to the sky, and therefore our measurement model is:
\begin{align}
M_j = T\otimes P_j + \epsilon_j,
\end{align}
where $\epsilon_j$ is Gaussian with mean $0$ and variance \begin{align}V[\epsilon_j] = B_j + F_jT\otimes P_j,\end{align}
where $B$ is the background noise, and $T\otimes P_j$ is the Poisson noise of the source.

In order to calculate the $S/N$ we refer to the fact
that $(S/N)^{2}$ is an additive quantity.
That is, for any two uncorrelated statistics $S_1,S_2$, we can build a statistic \begin{align}S_{1,2} = \alpha_1S_1 + \alpha_2S_2\end{align} such that the $S/N$ of $S_{1,2}$ satisfies \begin{align}(S/N)^2[S_{1,2}] = (S/N)^2[S_1] + (S/N)^2[S_2]\,.\end{align}
This fact is proven in Appendix~\ref{AppendixMaxSNR}.
Using this, all that we need to do in order to calculate the total $(S/N)^2$ of the optimal coaddition
of a source with flux $T_{0}$ in an ensemble of images
is to sum the $(S/N)^{2}$ over all the pixels in all the images.

Therefore, the $(S/N)^{2}$ for {\it flux} measurement (the source noise term in the variance is irrelevant for detection purposes) is given by:
\begin{align}
(S/N)^2 &= \sum_j\int_{-\infty}^{\infty}{\frac{T_0^2F_j^2P_j^2(x,y)}{V[M_j(x,y)]} dxdy} \\ \nonumber & = \sum_j\int_{-\infty}^{\infty}{\frac{T_0^2F_j^2P_j^2(x,y)}{T_0F_jP_j(x,y) + B_j} dxdy},
\label{eq:SNpsf_gen}
\end{align}

For the sake of signal to noise calculation, it is a good enough approximation to assume that the PSF is a two dimensional symmetric Gaussian. 
Rewriting the above with a Gaussian PSF, with width $s_j$, and replacing the $x$ and $y$ coordinates by the radial coordinate $r$
this can be written as
\begin{align} 
(S/N)^2 &= \\ \nonumber & =\sum_j\int_{0}^{\infty}{\frac{(T_0F_je^{-r^{2}/(2s_j^{2})}/[2\pi s_j^{2}])^{2}}{T_0F_je^{-r^{2}/(2s_j^{2})}/(2\pi s_j^{2}) + B_j}2\pi rdr}\,. \end{align}
This integral has an analytic solution given by:
\begin{align} \label{eq:SNpsf_gauss}
(S/N)^2 =  T_0\sum_jF_j - \sum_j2\pi s_j^{2}B_j\ln\left({1+\frac{F_jT_0}{2\pi s_j^{2}B_j}}\right)\,.
\end{align}


In the source-noise dominated case ($F_jT_0\gg2\pi s_j^{2}B_j$), 
the formula simplifies to
\begin{align}
(S/N)^{2} = T_0\sum_jF_j - \sum_j2\pi s_j^{2}B_j\left[\ln{(F_jT_0)} - \ln{(2\pi s_j^{2}B_j)}\right]\, .
\end{align}
In the background-noise dominated case ($F_jT_0\ll2\pi s_j^{2}B_j$), the $(S/N)^2$ becomes:
\begin{align}\label{eq:SNRFormulaMatchedFilter}
(S/N)^{2} = \sum_j{\frac{T_0^2F_j^{2}}{4\pi s_j^{2}B_j}}
\end{align}
We stress that this equation is correct only if the numbers $T_0$ and $B_j$ are in units of electrons.
Another remark is that Equation~\ref{eq:SNRFormulaMatchedFilter} is actually the exact (no approximation) $(S/N)^2$ for the detection of a source with flux $T_0$, because in this situation, the relevant variance should be exactly $V[M_j|\mathcal{H}_0]$.
Therefore, by isolating $T_0$ in Equation \ref{eq:SNRFormulaMatchedFilter} we can get a formula for the limiting magnitude for detection at a given $S/N$:
\begin{align}
T_0 = \sqrt{\frac{(S/N)^2}{\sum_j{\frac{F_j^2}{4\pi s_j^2B_j}}}}
\end{align}


%
%


\section{Robust coaddition}
\label{sec:WeightedMedian}
A drawback of the method presented so far is
that it uses the summation operator which is not resilient
against outliers.
Outliers can be a major source of noise and often require a special treatment.
The typical way outliers are dealt with is either using median coaddition, min/max rejection, or sigma clipping.

The summation in either Equation~\ref{eq:S_all} or \ref{eq:S_conv} includes weights, and in either case we may want to choose an operation which is more robust.
Because of the weights in Equations~\ref{eq:S_all} and \ref{eq:S_conv}, a simple median operation is not possible.
Instead, one can use the weighted median estimator to replace the weighted average.
The weighted median of a set of estimators
\begin{align}
\theta_j = \mu_j T + \epsilon_j
\end{align}
 is simply defined as
the 50\% weighted percentile of the normalized estimators $\frac{\theta_j}{\mu_j}$, where each of the estimators is given the weight of the inverse standard deviation\footnote{Inverse standard deviation and not inverse variance. A way to derive this is to define a symmetrical exponential loss function for the error in each image and find the value that minimizes the global loss function. This value is given analytically by the weighted median.}, $\frac{\mu_j}{\sigma_j}$. 
In a similar manner, one could apply sigma clipping to the estimators, removing estimators that deviate from the average estimated value of $T$ by more than 3 of its own standard deviation.

It is important to note, that these operations make sense only when we measure the flux of unblended objects using the matched filtered images\footnote{Performing median on regular images will yield weird artifacts when performed on images with different PSFs. Examples for such are seeing distribution dependent bias, flux dependent PSF, and the effective removal of any observation with a PSF which is different (better or worse) from the majority of the images.}.
Note that since our method uses the matched filter image as the estimator, sharp outliers like cosmic rays or hot pixels
will tend to reduce in significance (as
we are using the wrong filter for their detection).

Given the problems involved in using robust estimators, we would like to suggest a different approach, which is to identify and remove sharp outliers using image subtraction, given an artifact free image subtraction statistic. Such an image subtraction algorithm can be found in \cite{Sub}.
In the situation where a large number of images is coadded ($\gtorder10$),
any outlier that will influence the final measurement must be
easily visible in the subtraction image of any two images. Therefore, the detection and removal of any significant outliers prior to coaddition should be possible, and probably constitutes the most accurate approach for dealing with artifacts.
We will further discuss this in \cite{Sub}.

\section{Tests}
\label{sec:tests}

To verify that our method indeed works properly
and to try and identify possible problems
we tested it on simulated data (\S\ref{subsec:simulations})
and real data (\S\ref{subsec:real}).

\subsection{Simulated data}
\label{subsec:simulations}

The actual improvement of the coaddition technique
depends on the distribution of the observing conditions
(e.g., seeing, background and transparency).
Therefore, we use parameters that roughly mimic
typical observing conditions.
Obviously, for different surveys and use cases these parameters could vary significantly.

We repeated the following simulation 100 times: 
We simulated 20 $1024\times1024$\,pix$^{2}$ images, each with 2048 stars in
an equally spaced grid.
For simplification, we assumed all the images have equal transparency.
The seeing distribution was taken as uniform with full width at half maximum (FWHM)
between 1.5 and 6 pixels. The background distribution
was uniform between 500 to 1900\,e$^{-}$\,pix$^{-1}$.
The point spread functions were assumed to be circular two dimensional Gaussians.
We adjusted the brightness of the sources to be close to the detection limit in the coadd image (100 photons per image, see \S\ref{sec:TheoreticalSensitivity}).
The images were created aligned, so no further registration was applied.

Next, we summed the images using various coaddition schemes.
The techniques we tested are:
(i) Weighted summation of the images using the \cite{Annis2014}
scheme (see Equation~\ref{eq:Annis});
(ii) Weighted summation of the images using the \cite{Jiang2014}
method (see Equation~\ref{eq:Annis});
(iii) Our method.
We note that unlike \cite{Annis2014} or \cite{Jiang2014}
we did~not remove any images\footnote{Given the fact that weights (if calculated correctly)
should take care of low quality images,
we regard this step as damaging rather than beneficial.}.
To evaluate the detection $S/N$ of each method, we averaged the $S/N$ of all sources at their correct positions in the coaddition image (using Eq. \ref{eq:S_sig}). Given the number of sources simulated this average is accurate to the level of about $1\%$.
We then calculated, using the specific seeing and brightness parameters of the images, the theoretically maximal $S/N$ for source detection (Eq. \ref{eq:SNRFormulaMatchedFilter}).

\begin{figure}
\centering
\includegraphics[width = 85mm]{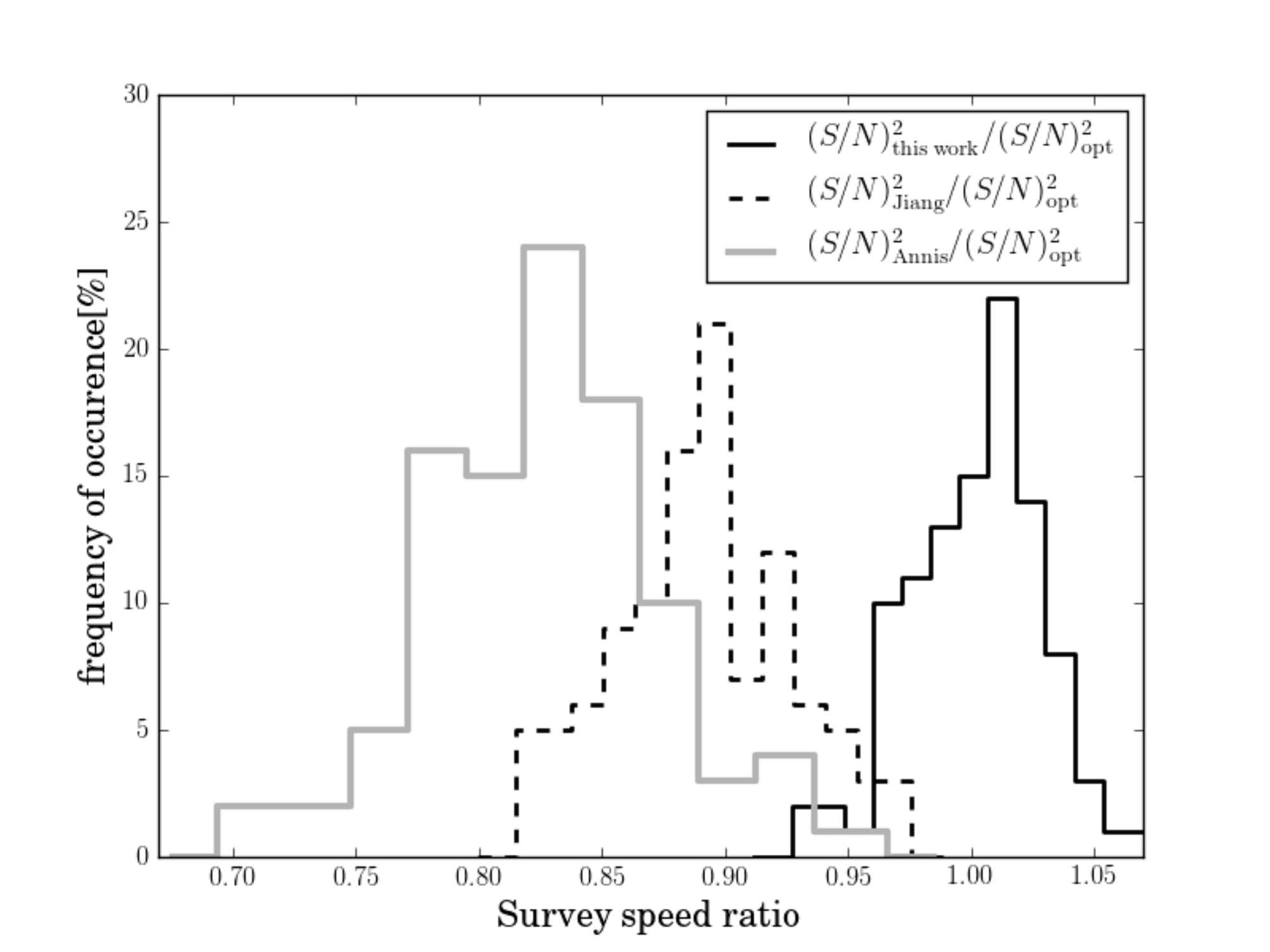}
\caption{Signal to noise distributions in the simulated images. For each coaddition method, and for each simulation the average of the achieved $(S/N)^2$ for source detection is calculated. The distributions of these averages divided by the theoretically calculated maximum $(S/N)^2$ are shown for each coaddition method. \label{fig:SNR_distributions}}
\end{figure}

In Figure \ref{fig:SNR_distributions}, we plot the distribution of the ratio of the obtained $S/N$ with the various methods and the theoretically calculated $S/N$.
It is clear that the method presented in this paper achieves the maximum possible $S/N$, and that the symmetric scatter around it's value is the measurement error of the achieved $S/N$.
It is further clear that for this specific case, our method provides a $\sim 5\%$ higher $S/N$ than the method of \cite{Jiang2014}, and a $\sim 10\%$ higher $S/N$ than \cite{Annis2014}.
These translate to a survey speed increase of about 10\% over \cite{Jiang2014}, and 20\% over \cite{Annis2014}.  

\subsection{Real data}
\label{subsec:real}

We performed several comparisons
of our method with other techniques,
using real data.
Our test data originates from the Palomar Transient Factory
(PTF\footnote{http://www.ptf.caltech.edu/iptf}; Law et al. 2009; Rau et al. 2009)
data release 2.
The images were obtained under various atmospheric conditions.
All the images were reduced using the PTF/IPAC pipeline
(Laher et al. 2014),
and were photometrically calibrated (Ofek et al. 2012).

\begin{figure*}
\centerline{\includegraphics[width=16cm]{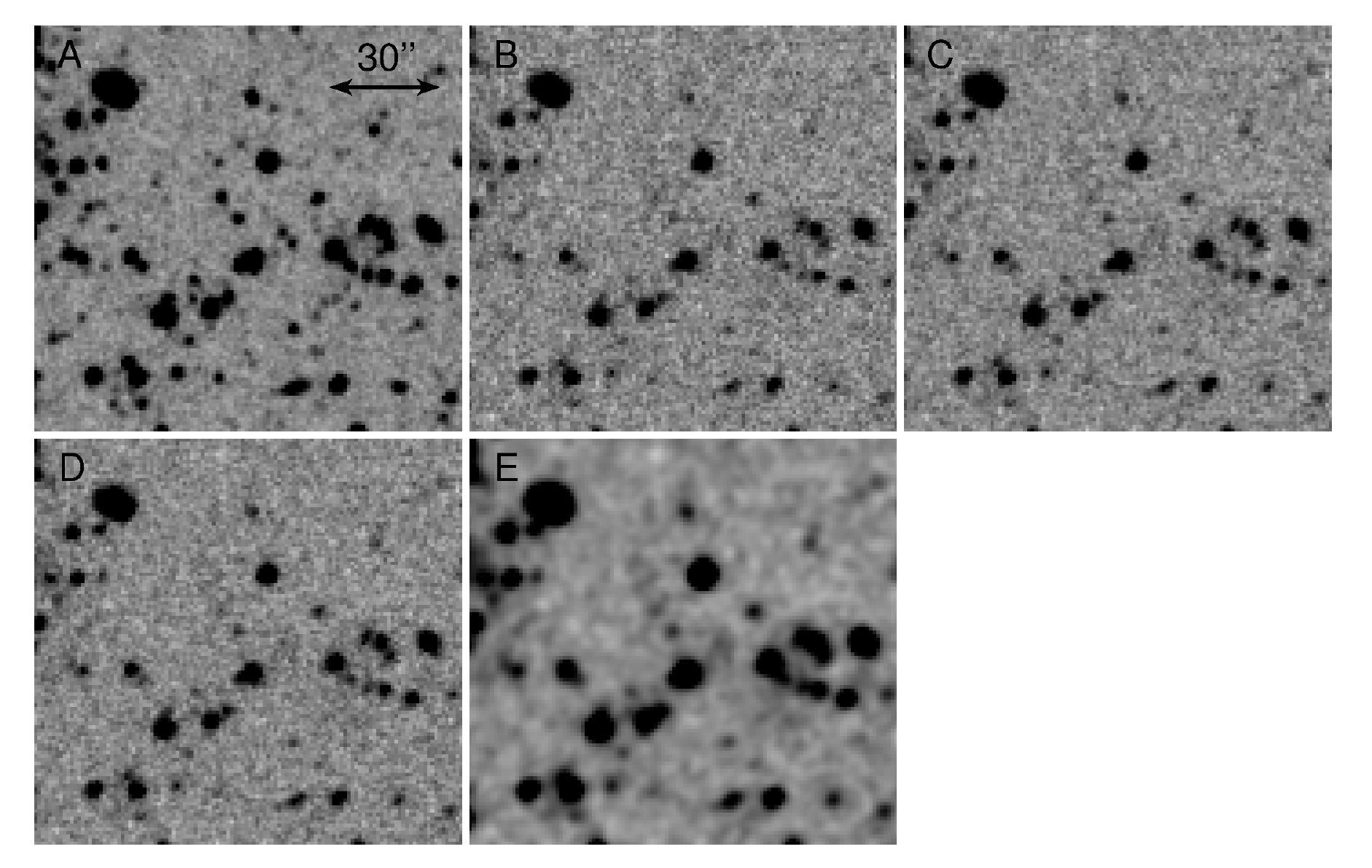}}
\caption{Small cutouts of coadded images based on data-set 3.
A - deep coadd image; B - no weights no filter; C - Coaddition with the Annis et al. (2014) weights;
D - Coaddition with the Jiang et al. (2014) weights; E - optimal coaddition for source detection.
10\% variations in information content can~not be detected by eye, but quantitative comparison
shows that the optimal coaddition for source detection provide improved $S/N$ for sources,
and hence larger survey speed.
\label{fig:CoaddComp}}
\end{figure*}

We used four sets of images. For each set we prepared a deep reference
image.
The reference image is based on coaddition of a subset of the best images
using the Annis et al. (2014) weights and no filtering.
We also selected a subset of images which we coadded using the various techniques
including:
equal weights, Annis et al. (2014) weighting,
Jiang et al. (2014) weighting, our optimal-detection coaddition and the optimal coaddition method described in paper~II.
In order to minimize non-linear registration effects we used
a section of $1024\times1024$\,pixels near the center of each field.
The four sets of images, along with their selection criteria,
are listed in Table~\ref{tab:Selection}.
Small cutouts of the coadd images are presented in Figure \ref{fig:CoaddComp}.
We note that it is very difficult to spot the fine differences between
these images by eye and quantitative tests are required.

\begin{deluxetable*}{llllll}
\tablecolumns{5}
\tablewidth{0pt}
\tablecaption{Sets of images used in the comparison}
\tablehead{
\colhead{Set}         &
\colhead{Field}       &
\colhead{CCD}         &
\colhead{\#im}         &
\colhead{Type}        &
\colhead{Criteria}     \\
\colhead{}            &
\colhead{}            &
\colhead{}            &
\colhead{}            &
\colhead{}            &
\colhead{}
}
\startdata
1 &  100031 & 6 & 425  & Ref.   & Variance$<$1000\,e$^{-}$ \& FWHM$<$4'' \& $>10$ PSF stars \\
  &         &   &  45  & coadd  & All images taken on Oct, Nov, Dec 2012 \& $>10$ PSF stars \\
\hline
2 &  100031 & 6 & 425  & Ref.   & Variance$<$1000\,e$^{-}$ \& FWHM$<$4'' \& $>10$ PSF stars \\
  &         &   &  48  & coadd  & All images taken on the first 9 days of each Month in 2011 \& $>10$ PSF stars \\
\hline
3 &  100031 & 4 & 263  & Ref.   & Variance$<$1000\,e$^{-}$ \& FWHM$<$4'' \& $>10$ PSF stars \\
  &         &   &  39  & coadd  & All images taken on Oct, Nov, Dec 2012 \& $>10$ PSF stars \\
\hline
4 &  100031 & 4 & 263  & Ref.   & Variance$<$1000\,e$^{-}$ \& FWHM$<$4'' \& $>10$ PSF stars \\
  &         &   &  17  & coadd  & All images taken on the first 9 days of each Month in 2011 \& $>10$ PSF stars 
\enddata
\tablecomments{Selection criteria for reference images and coadd images
in the various sets used for testing the coaddition methods.
Field indicate the PTF field ID,
while CCDID is the CCD number in the mosaic (see Law et al. 2009; Laher et al. 2014).
Type is either Ref. for the deep reference image, or Coadd for the subset
we coadd using the various methods.
Note that field 100031 is in the vicinity of M51.
\label{tab:Selection}}
\end{deluxetable*}

We multiply each image by its CCD gain (to work in electron units).
The images were coadded using {\tt sim\_coadd.m} (Ofek 2014).
This function first registers the images,
using {\tt sim\_align\_shift.m},
subtracts the background from all the images
({\tt sim\_back\_std.m}),
calculates the weights (for the various methods, {\tt weights4coadd.m}),
and estimates the PSF for each image ({\tt build\_psf.m}),
and finally coadds the images.
The program optionally filters each image with its PSF prior
to coaddition ({\tt sim\_filter.m}).
All these functions are available\footnote{http://webhome.weizmann.ac.il/home/eofek/matlab/} as part of
the astronomy and astrophysics toolbox for MATLAB \citep{Ofek14} (see additional details in \S\ref{code}).
We note that the coaddition technique described in paper~II,
is implemented in {\tt sim\_coadd\_proper.m}.

In order to compare between the various techniques,
we first run our source extraction code {\tt mextractor.m}.
Like {\tt SExtractor} \citep{Bertin96}, this code uses 
linear matched filter to search for sources.
However, while {\tt SExtractor} thresholds the filtered image
relative to the standard deviation of the {\it un}-filtered image,
our code does the thresholding relative to the
filtered image.
This small, but important, difference means that our thresholding
is always done in units of standard deviations, so images
with different PSFs can be compared easily.
We set the detection threshold to $4\sigma$.
We note that the coadded images were always filtered using
their PSF.

We matched the sources found in each coadded image,
against sources found in the deep reference image.
Table~\ref{tab:Comp} lists the number of sources detected in each image ($N_{{\rm d}}$),
the number of sources in the reference that are matched in the coadd ($N_{{\rm r}}$),
the number of sources in the reference that are un-matched in the coadd image ($N_{{\rm u}}$),
and the number of false detections in the coadd ($N_{{\rm f}}$).
Our method consistently finds more (3\% to 12\%) real sources than other weighted-summation methods. We note that the slight increase in the number
of false sources using our method is due to the effective better PSF that our method has
relative to regular weighted addition (after matched-filtering; see paper~II).
An image with better seeing, once matched filtered, has a larger number of uncorrelated scores,
and therefore an increased amount of sources with score that is larger than some threshold.
This effect is reproduced in simulations, and could be easily accounted for in the survey design by
slightly increasing the threshold above which a source is declared statistically significant.
It is further important to note that even if we fix the number of false positives to some constant,
our method still detects more sources than the other techniques.
In Table~\ref{tab:Comp} we also list the results based on the method presented in paper~II.
For source detection, the technique described in paper~II is identical (up to small numerical errors) to
the method discussed in this paper. However, the method described in paper~II
has several important advantages, and hence should be used whenever adequate.
\begin{deluxetable}{lllllll}
\tablecolumns{7}
\tablewidth{0pt}
\tablecaption{Comparison between coaddition methods}
\tablehead{
\colhead{Set}               &
\colhead{Method}            &
\colhead{$I_{{\rm r}}$}            &
\colhead{$N_{{\rm d}}$}     &
\colhead{$N_{{\rm r}}$}     &
\colhead{$N_{{\rm u}}$}     &
\colhead{$N_{{\rm f}}$}\\
\colhead{}            &
\colhead{}            &
\colhead{}            &
\colhead{}            &
\colhead{}            &
\colhead{}            &
\colhead{}
}
\startdata
1 & Deep            &      0  &       2433    &   \nodata  &    \nodata   &   \nodata \\
  & Equal weights   & 0.65    &       1136    &     1087   &      1346    &       49  \\
  & \cite{Annis2014}&0.91     &       1332    &     1255   &      1178    &       77  \\
  & \cite{Jiang2014}&0.89     &       1299    &     1229   &      1204    &       70  \\
  & This work       &      1  &       1417    &     1324   &      1109    &       93  \\
  & paper~II        &      1  &       1419    &     1325   &      1108    &       94  \\
\hline
2 & Deep            &      0  &       2433    &   \nodata  &    \nodata   &   \nodata \\
  & Equal weights   &0.59     &       1183    &     1085   &      1348    &       98  \\
  & \cite{Annis2014}&0.93     &       1481    &     1343   &      1090    &      138  \\
  & \cite{Jiang2014}&0.90     &       1421    &     1301   &      1132    &      120  \\
  & This work       &      1  &       1541    &     1390   &      1043    &      151  \\
  & paper~II        &      1  &       1542    &     1391   &      1042    &      151  \\
\hline
3 & Deep            &      0  &       4062    &   \nodata  &    \nodata   &   \nodata \\
  & Equal weights   &0.53     &       1645    &     1539   &      2523    &      106  \\
  & \cite{Annis2014}&0.84     &       1955    &     1796   &      2266    &      159  \\
  & \cite{Jiang2014}&0.81     &       1912    &     1759   &      2303    &      153  \\
  & This work       &      1  &       2206    &     1976   &      2086    &      230  \\
  & paper~II        &      1  &       2206    &     1976   &      2086    &      230  \\
\hline
4 & Deep            &      0  &       4062    &   \nodata  &    \nodata   &   \nodata \\
  & Equal weights   &0.83     &       2319    &     2007   &      2055    &      312  \\
  & \cite{Annis2014}&0.82     &       2144    &     1981   &      2081    &      163  \\
  & \cite{Jiang2014}&0.92     &       2272    &     2062   &      2000    &      210  \\
  & This work       &      1  &       2401    &     2125   &      1937    &      276  \\
  & paper~II        &0.99     &       2400    &     2124   &      1938    &      276  
\enddata
\tablecomments{Quantitive comparison between images coadd using various technique
matched against a deep reference image.
The four blocks, separated by horizontal lines correspond to the four data sets in Table~\ref{tab:Selection}.
$I_{{\rm r}}$ is the approximate survey speed ratio between the image compared and our optimal coaddition method (see text for details).
$N_{{\rm d}}$ is the number of detected sources to detection threshold of 4$\sigma$.
$N_{{\rm r}}$ is the number of real sources (i.e., detected sources which have a counterpart
within $3''$ in the deep reference image.
$N_{{\rm u}}$ is the number of real undetected sources (i.e., sources detected in the reference which are not detected in the coadd image).
$N_{{\rm f}} = N_{d}-N_{r}$ is the number of false detections.
\label{tab:Comp}}
\end{deluxetable}

\begin{figure}
\centerline{\includegraphics[width=8.5cm]{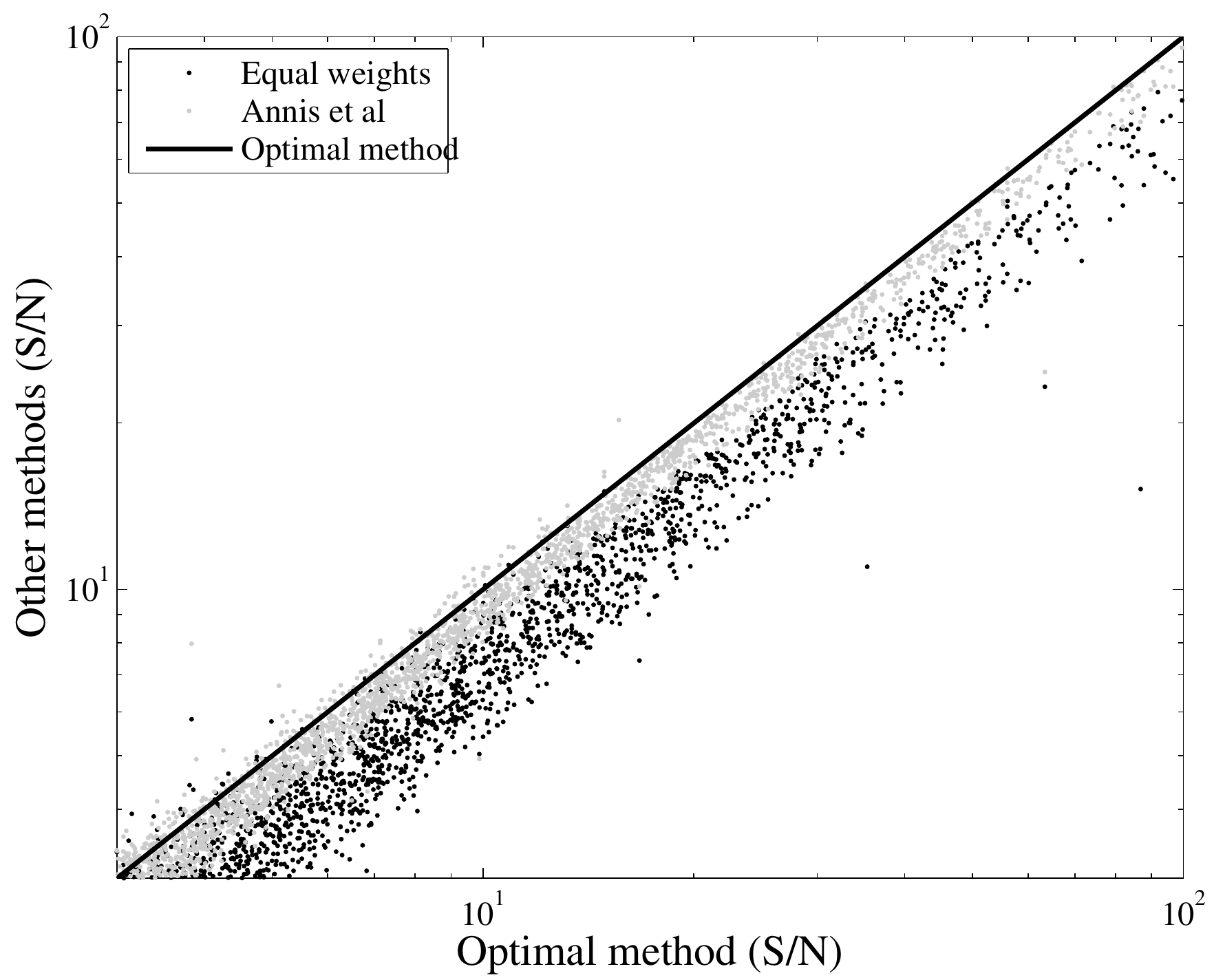}}
\caption{The ratio between detection $S/N$ of individual sources
in a coadded image relative to the optimal coaddition.
Our method is represented by the 1:1 line, while
the \cite{Annis2014} and un-weighted coaddition
are gray and black dots, respectively.
This example is based on the third data set.
\label{fig:SNcomp}}
\end{figure}

Next, we compared the detection
$(S/N)^{2}$ of the sources in the various images\footnote{Note that the detection $S/N$ is defined without the source-noise term
in the denominator (see Eq. \ref{eq:SNRFormulaMatchedFilter}).}.
The reason for using $(S/N)^{2}$ is that it is an additive quantity, that is proportional to the survey speed and to the detection information.
In order to do the comparison,
we run {\tt mextractor.m} again in forced-detection mode.
In this mode, we provide the code with a list of positions
of the sources detected in the deep reference image,
and the code measures the detection significance and $S/N$ at these locations.
For each one of the different coaddition methods on the subset of images we calculated the following quantity:
For each star detected in the reference image, we divide its $S/N$
measured\footnote{The $S/N$ is measured by filtering the image with its PSF,
normalizing the filtered image by its own standard deviation, and reading the value
at the local maximum that corresponds to the source.}
in a coaddition image by its S/N measured in our optimally coadded image.
Then, we sum the squares of these ratios, and calculate their median.
This roughly gives the survey speed in each coadd image,
relative to our coaddition method,
and is listed in Table~\ref{tab:Comp} ($I_{{\rm r}}$).
In these specific examples, our method provide a factor of 17\% to 47\% improvement over un-weighted coaddition,
and 8\% to 18\% improvement over the weighted addition methods.
We note that this is sensitive to the exact seeing distribution of the observations
and we estimate that the typical improvement will be between a few percents to 25\%,
relative to \cite{Annis2014}.
Finally, in Figure~\ref{fig:SNcomp} we present, for data set 3, the relative ratio between
detection $S/N$ of individual sources in the coadded images relative to
the detection $S/N$ in the optimal coadd image.
Again, our coaddition provides better $S/N$ both at the faint and bright ends.
We note that using the techniques described in \S\ref{sec:SourcePhotometry}, the flux measurement $S/N$ in the bright end can be further improved.

\section{Code and implantation details}
\label{code}
The formulae we present in this paper are straight forward to code.
However, most of the attention in the implementation should be given to pre-processing steps and
measurements of the properties of the images. This includes estimating the background mean, the
variance, the PSF and cosmic ray removal.

Code that performs the coaddition technique suggested in this paper is available as part of the Astronomy and Astrophysics toolbox for
MATLAB \citep{Ofek14}.
This package can be used for all the image processing steps,
including the de-bias, flat field correction and cosmic-ray removal.
However, here we cover only the functions that are closely related to the coaddition step.

Table~\ref{tab:Fun} lists the high-level functions related to coaddition that
we provide in the Astronomy and Astrophysics toolbox for
MATLAB \citep{Ofek14}.
Many other low-level functions are documented in the code.
These functions are under constant improvement, and we expect
that versions with better performances will be available in the near future.
Each function has a detailed help section with examples.
Furthermore, a manual of the Astronomy and Astrophysics toolbox for
MATLAB is also available.

We note that usual implementation details like
registration, resampling to the same grid,
measuring the background and variance levels, and measuring/interpolating
the PSF, as always, require attention.
However, the attention to the details required by this method is not different than that required for the successful application of other methods.
%
Below we comment on several important implementation details:

{\bf Background and variance estimation}:
The background and variance in real wide field of view
astronomical images cannot be treated as constants over the entire field of view.
Therefore, we suggest to estimate them locally and interpolate.
To estimate the background and variance one needs to make sure that the
estimators are not biased by stars or small galaxies.
Our suggestion is to fit a Gaussian to the histogram of the image pixels
in small regions\footnote{We are currently using $256\times256$~arcsec$^{2}$ blocks.},
and to reject from the fitting process pixels with high
values (e.g., the upper 10-percentile of pixel values).

{\bf Estimating the transparency}:
The transparency $F_{j}$ of each image
is simply its flux-based photometric zero point\footnote{See appendix A of \cite{Ofek11} for a method to calculate this zero-points.}.
However, one has to make sure that these zero points are measured
using PSF photometry rather than aperture photometry,
otherwise the zero-points may depend on the seeing.

{\bf Estimating the PSF}:
Among the complications that may affect the PSF measurement
are pixelization, interpolation and resampling.
Furthermore, the PSF is likely not constant spatially and it also
may change with intensity due to charge self repulsion.
This specifically may lead to the brighter-fatter effect (e.g., Walter 2015).
We note that the fact that one needs to estimate the PSF in order to run
this method should not be viewed as a drawback, as any decent method
that finds sources in the image requires this step anyway.


%
\begin{deluxetable*}{lll}
\tablecolumns{2}
\tablewidth{0pt}
\tablecaption{High-level functions relevant for coaddition}
\tablehead{
\colhead{Name}               &
\colhead{Description  }    \\
\colhead{}           &
\colhead{}
}
\startdata
{\tt sim\_coadd.m}             & Coadd a list of images, using various weighting schemes.\\
                               & The function also allows for filtering the images prior to the coaddition.\\
                               & The function can also align the images, calculate the weights and PSFs.\\
{\tt sim\_coadd\_proper.m}     & Proper coaddition of images (see \citep{Coad2}). \\
                               & The function can also align the images, calculate the weights and PSFs.\\
\hline
{\tt sim\_align\_shift.m}      & Register a set of images against a reference image.\\
                               & The function assumes the images can be registered \\
                               & using an arbitrary large shift, but only a small rotation term.\\
{\tt psf\_builder.m}           & Construct a PSF template by re-sampling the pixels around \\
                               & selected bright/isolated stars.\\
{\tt weights4coadd.m}          & Calculate parameters required for calculation of weights for \\
                               & coaddition. Including the background, its variance, estimate of the \\
                               & flux-based zero points (i.e., transparency), and measure the PSF.\\
{\tt sim\_back\_std.m}         & Estimate the spatialy-dependent background and variance of images.
\enddata
\tablecomments{High-level functions relevant for coaddition, which are part of the
Astronomy and Astrophysics toolbox for MATLAB (Ofek 2014).}
\label{tab:Fun}
\end{deluxetable*}

\section{Summary}
\label{sec:Disc}

%
%
%
%
%

We argue that popular image coaddition methods do~not achieve the maximal $S/N$ possible for source detection and photometry.
We derive the optimal statistic for source detection and photometry under the assumptions that the noise is approximately Gaussian
and that the target is well separated from other bright sources.
We show that the optimal way to coadd images for source detection is by filtering
(i.e., cross-correlating) each image with its PSF, and then
sum with weights.
This method is summarized by Equation~\ref{eq:S_conv} (or equivalently, Eq. \ref{eq:S_hat}).
In order to find sources we need to find local maxima in the calculated score map.
The significance of these sources in units of standard deviations is given by Equation~\ref{eq:S_sig}.
We note that the computational requirements of applying the method are low, as the cross-correlation operation
can be computed using the fast Fourier transform (FFT)\footnote{The run time of FFT on an image is typically an order of magnitude faster than reading/writing an image from the hard disk.}.
We also derive optimal statistics for PSF photometry (Eq.~\ref{eq:T_PSF_phot})
and aperture photometry (Eq.~\ref{eq:Theta_aper_phot}) for an ensemble of images.
Finally we derive a formula to calculate the $S/N$
for PSF photometry in an ensemble of images with a symmetric
Gaussian PSF (Eq.~\ref{eq:SNpsf_gauss}).
For statistical tasks other than source detection or photometric measurements, we refer readers to paper~II in this series, in which we provide a coaddition method that is optimal for {\it any} statistical measurement or decision, under the more restrictive (but common) assumption of background dominated noise. 

We demonstrate our coaddition method for source detection
on simulated images as well as on real images.
Our method increases the survey speed (for faint source detection
and photometry) by between a few percents to 25\% over traditional methods, both in theory and in practice.

\acknowledgments

We thank Avishay Gal-Yam, Assaf Horesh, Frank Masci, William Newman, and Ora Zackay for discussions.
This paper is based on observations obtained with the
Samuel Oschin Telescope as part of the Palomar Transient Factory
project, a scientific collaboration between the
California Institute of Technology,
Columbia University,
Las Cumbres Observatory,
the Lawrence Berkeley National Laboratory,
the National Energy Research Scientific Computing Center,
the University of Oxford, and the Weizmann Institute of Science.
B.Z. is grateful for receiving the Clore fellowship.
E.O.O. is incumbent of
the Arye Dissentshik career development chair and
is grateful for support by
grants from the 
Willner Family Leadership Institute
Ilan Gluzman (Secaucus NJ),
Israel Science Foundation,
Minerva, Weizmann-UK,
and the I-Core program by the Israeli Committee for Planning
and Budgeting and the Israel Science Foundation (ISF).

\appendix
\section{Weighted addition of independent random variables for hypothesis testing}\label{AppendixProperAddition}
Here, we will show that the same statistic that was derived in \S\ref{subsec:ExampleGaussians} as an exact solution for detection of an attenuated signal in the presence of varying Gaussian noise, is the solution of the much more robust question -- maximizing the $S/N$ of the weighted addition of independent random variables. 
Given a set of statistical variables, $X_j$ with two hypotheses $\mathcal{H}_0$ and $\mathcal{H}_1$, with expectencies:\footnote{We note that even if $E[X_j|\mathcal{H}_0]= z_j$, it is always possible to transfer the problem to hypothesis testing on $Y_j = X_j-z_j$, in which case the expectancy given the null hypothesis is again zero.} \begin{align}
E[X_j|\mathcal{H}_0] = 0\,,\quad E[X_j|\mathcal{H}_1] = \mu_j\,.
\end{align}
Further, we assume that the variance of $X_j$ is equal under the assumption of both hypotheses. i.e,
\begin{align}
V[X_j|\mathcal{H}_0] = V[X_j|\mathcal{H}_1] = \sigma_j^2\,.
\end{align}

If we know the exact distributions of the variables $X_j$ given both hypotheses, we can use the Neyman-Pearson lemma \citep{NeymanPearsonLemma}, which states that the optimal test statistic for the decision between $\mathcal{H}_0$ and $\mathcal{H}_1$ is the log-likelihood ratio test. Here, we want to construct a test statistic that we can apply even when the exact distributions are unknown, but their first and second moments are known.

Given the fact that all the statistics are assumed to be independent, it is reasonable to assume that the correct way to combine the variables is via a linear sum (if the distributions are known we can simply sum the difference in the log of the probablity of observing $X_j$ for each hypothesis).
That is, we are interested in a linear statistic of the form:
\begin{align}S = \sum_j{\beta_jX_j}\,.\label{Ap:eq:s}\end{align}

If the set of variables we have is large, so that the distribution of a weighted sum of the variables can be approximated by a Gaussian distribution for both hypotheses, then the only important properties of the resulting sum are the expectancy of $S$ given both hypotheses, and the variance of $S$ given both hypotheses. Moreover, if the statistic has equal variance under both hypotheses, then there is one number that characterizes our ability to distinguish between the hypotheses, which is the squared signal-to-noise ratio:
\begin{align}\label{Ap:eq:SNR}
\left(S/N\right)^2[S] \equiv \frac{|E[S|\mathcal{H}_0] - E[S|\mathcal{H}_1]|^2}{V[S]}\,.
\end{align} 
Thus, we want to find the linear combination that maximizes the $S/N$.
i.e., find a set of weights $\beta_j$ such that the score in Equation~\ref{Ap:eq:s} will have maximal $S/N$ (Equation~\ref{Ap:eq:SNR}).

We assume that $X_j$, $\mu_j$, and $\beta_j$ are complex variables
(this will become useful in future papers in the series). Substituting \{0,$\mu_j$\} as the expectancies of $X_j$ given the two hypotheses, we get:
\begin{align}
|E[S|\mathcal{H}_0] - E[S|\mathcal{H}_1]|^2 = |\sum_j{\beta_j\mu_j}|^2 \equiv |\Delta E[S]|^2\,.
\end{align}
Given the hypothesis $\mathcal{H}_\alpha$, where $\alpha \in \{0,1\}$, the variance of $S$ is:
\begin{align}V[S|\mathcal{H}_\alpha] = \sum_j{|\beta_j|^2\sigma_j^2}\,.\end{align}
In order to maximize the squared $S/N$ with respect to $\beta_j$, we equalize all the partial derivatives to zero: \begin{align}\frac{\partial \left(\left(S/N\right)^2[S]\right)}{\partial \beta_j} = 0\,, \end{align}
which gives us:
\begin{align}
0 = 2\frac{\partial \Delta E(S)}{\partial \beta_j}\frac{\Delta E}{V} - \frac{\partial V(S)}{\partial \beta_j}\frac{\Delta E^2}{V^2} = 2\overline{\mu_j}\frac{\Delta E}{V} - 2\beta_j\sigma_j^2\frac{\Delta E^2}{V^2}\,,
\end{align}
where the $\overline{\raisebox{1.5mm}{} \quad}$ accent denotes complex conjugation.
Simplifying, we get the equation:
\begin{align}
\beta_j=\frac{\overline{\mu_j}}{\sigma_j^2}\frac{V}{\Delta E}\,.
\end{align}
Noticing that multiplying all the weights $\beta_j$ by a constant factor does not change the $S/N$, we can simplify further and get:
\begin{align}
\beta_j=\frac{\overline{\mu_j}}{\sigma_j^2} = \frac{\overline{E[X_j|\mathcal{H}_1] - E[X_j|\mathcal{H}_0]}}{V_j}\,.
\end{align}
Thus, we now have a general formula for weighted addition of random variables.

We note that if the exact distributions of the random variables are known, then the optimal combination is the log-likelihood ratio test statistic. This statistic could be composed of some function of the variables themselves that is not linear. This means that the best score will be of the form \begin{align}\sum_j{f_j(x_j)}\,.\end{align} What we have derived in this appendix, is the best linear approximation to this score, that could be derived under much less exact knowledge on the variables themselves.


\section{Weighted addition of independent estimators}\label{AppendixProperAdditionEstimators}
Here, we will show that the same statistic that was derived in \S\ref{subsec:ExampleGaussians} and Appendix \ref{AppendixProperAddition} is the solution (up to a multiplicative constant) of another general question:
What is the maximum $S/N$ linear combination of independent estimators?

Given a set of statistics $\theta_j$ such that 
\begin{align}
\theta_j = \mu_j T + \epsilon_j\,,
\end{align}
where $\mu_j$ are known complex variables, \begin{align}E[\epsilon_j] = 0\,, \quad V[\epsilon_j] = \sigma_j^2\,.\end{align}
We want to find the best weighted linear combination estimator 
\begin{align}\theta = \sum_j{\beta_j\theta_j}\,,\end{align} such that $E[\theta] = T\,,$ and $V[\theta]$ is minimal.
Calculating the expectancy of $\theta$ we get:
\begin{align}E[\theta] = \sum_j{\beta_j\mu_jT}\,,\end{align}
while the variance of $\theta$ is:
\begin{align}V[\theta] = \sum_j{|\beta_j|^2\sigma_j^2}\,.\end{align}
Maximizing the $(S/N)^2$ of the estimator: \begin{align}
(S/N)^2[\theta] \equiv \frac{|E[\theta]|^2}{V[\theta]}\,,
\end{align} is equivalent to minimizing the variance of $\theta$ with respect to a fixed $E[\theta]$.
Moreover, $(S/N)^2[\theta]$ is invariant to scalar multiplication of $\theta$ because:
\begin{align}
(S/N)^2[\beta\theta] = \frac{|E[\beta\theta]|^2}{V[\beta\theta]} = \frac{|\beta|^2|E[\theta]|^2}{|\beta|^2V[\theta]} =(S/N)^2[\theta]\,.
\end{align}
Therefore, we can maximize $(S/N)^2[\theta]$ with respect to all $\beta_j$. After normalization the minimum variance estimator is: \begin{align}\tilde{\theta} = \frac{\theta}{\sum_j{\beta_j\mu_j}}\,.\end{align}

To maximize $(S/N)^2[\theta]$ with respect to $\beta_j$, we equalize all the partial derivatives to zero: \begin{align}\frac{\partial (S/N)^2[\theta]}{\partial \beta_j} = 0\,, \end{align}
which gives us:
\begin{align}
0 = 2\frac{\partial E[\theta]}{\partial \beta_j}\frac{E[\theta]}{V[\theta]} - \frac{\partial V(\theta)}{\partial \beta_j}\frac{E[\theta]}{V[\theta]^2} = 2\overline{\mu_j}\frac{E[\theta]}{V[\theta]} - 2\beta_j\sigma_j^2\frac{E[\theta]}{V[\theta]^2}\,,
\end{align}
Where the accent $\overline{\strut \quad}$ denotes complex conjugation.
Simplifying, we get the equation:
\begin{align}
\beta_j=\frac{\overline{\mu_j}}{\sigma_j^2}\frac{V[\theta]}{E[\theta]}\,.
\end{align}
Because $(S/N)^2[\theta]$ is invariant under multiplication by a  constant factor, we get:
\begin{align}
\beta_j=\frac{\overline{\mu_j}}{\sigma_j^2}\,.
\end{align}
Adjusting the estimator to match our first requirement that $E[\theta] = T$ we can finally write:

\begin{align}
\tilde{\theta}=\frac{\sum_j{\frac{\overline{\mu_j}}{\sigma_j^2}X_j}}{\sum_j{\frac{|\mu_j|^2}{\sigma_j^2}}}\,.
\end{align}

\section{$S/N$ of the addition of optimally weighted statistics }\label{AppendixMaxSNR}
In Appendix \ref{AppendixProperAddition} we find the best way to add statistics even when we do~not know their exact distributions . Here we would like to derive a closed formula for the $S/N$ of the resulting statistic.
Denote by \begin{align} \left(S/N\right)^2[X_j] = \frac{|\Delta E[X_j]|^2}{V[X_j]}\,.\end{align}
Writing the expression for the statistic $S$, the maximum $(S/N)$ weighted addition of the variables defined in Appendix \ref{AppendixProperAddition}:
\begin{align}
S = \sum_j{\frac{\overline{\Delta E[X_j]}\,X_j}{V[X_j]}}\,.
\end{align}
Calculating the expected difference in $S$ given the two different hypotheses we get:
\begin{align}
\Delta E[S] = E[S|\mathcal{H}_1] - E[S|\mathcal{H}_0] = \sum_j{\frac{|\Delta E[X_j]|^2}{V[X_j]}}\,,
\end{align}
while the variance of $S$ is:
\begin{align}
V[S] = \sum_j{\frac{|\Delta E[X_j]|^2}{V[X_j]^2}V[X_j]} = \sum_j{\frac{|\Delta E[X_j]|^2}{V[X_j]}}\,.
\end{align}
Therefore, the squared $S/N$ of $S$ is simply given by:
\begin{align}
\left(S/N\right)^2[S] = \frac{\left(\sum_j{\frac{|\Delta E[X_j]|^2}{V[X_j]}}\right)^2}{\sum_j{\frac{|\Delta E[X_j]|^2}{V[X_j]}}} = \sum_j{\frac{|\Delta E[X_j]|^2}{V[X_j]}} = \sum_j{\left(S/N\right)^2[X_j]}\,.
\end{align}

This means that the squared $S/N$ of a statistic which is the optimal weighted addition of a set of statistics is the sum of squared $S/N$'s of the individual statistics.
This property now allows us to easily estimate the sensitivity of observations ahead of time, with an intuitive closed formula.
This (almost) same calculation applies also for optimal weighted addition of estimators, and we omit it to prevent cumbersome repetitions. 

\section{How much survey speed is lost when we use the wrong PSF?}
\label{Ap:InfoPSFlost}

Here we derive an analytic solution to the question how much sensitivity is lost in the detection process when the wrong
linear matched filter is used.
Following previous sections, we define the measured image by:
\begin{align}
M = T\otimes P + \epsilon,
\end{align}
and assume that the source we are trying to detect is well separated from other sources. We also assume that the source we are looking for is faint relative to the background, leading to a spatially invariant noise variance:
\begin{equation}
V[\epsilon(x)] = \sigma^2\,.
\end{equation}
For simplification, we assume that the PSF, $P$, can be approximated as a symmetric Gaussian.
We define $s_{r}$ as the width of the real PSF $P$,
while $s_{u}$ is the width of the Gaussian used for match filtering.
Denoting by $w_x(s_u)$ the weight for the pixel $x$, when using a kernel with width $s_u$. Now, we can calculate the $(S/N)^2$ of the statistic \begin{align}W(s_{u}) = \sum_x{w_x(s_u)M(x)}\,.\end{align} 
by:
\begin{align}
(S/N)^2[W(s_u)]  = \frac{E[W(s_u)]^2}{V[W(s_u)]} =  \frac{(\sum_{x}{w_{x}(s_u)E[M(x)]})^{2}}{\sum_{x}{w_{x}^{2}(s_u)V[M(x)]}}\,.
\end{align}
Substituting the PSF and the match filtering kernel with the relevant Gaussians, and approximating the sums with integrals, we get: 
\begin{align}
    (S/N)^2[W(s_u)] &= \frac{T_0^2\frac{1}{ (2\pi)^{2}s_{u}^{2}s_{r}^{2}}\int{e^{-\vert x\vert^{2}/(2s_{u}^{2})} e^{-\vert x\vert^{2}/(2s_{r}^{2})} dx_{1}dx_{2}}  }{\sigma^2\frac{1}{ (2\pi)^{2}s_{u}^{4}}\int{e^{-\vert x\vert^{2}/(s_{u}^{2}) }  dx_{1}dx_{2}}   } 
   \\&= \frac{T_0^2}{\pi\sigma^2} \frac{s_{u}^{2}}{(s_{u}^{2}+s_{r}^{2} )^{2}} = \frac{T_0^2}{\pi\sigma^2s_r^2} \frac{s_{u}^{2}s_{r}^{2}}{(s_{u}^{2}+s_{r}^{2} )^{2}}\,,
\label{eq:InfoPSFlost}
\end{align}
where $\vert x\vert^{2} = x_{1}^{2} + x_{2}^{2}$.

The maximum of this function with respect to a given value of $s_r$ is obtained at $s_u = s_r$.
When substituting $s_u = ks_r$, and dividing the $(S/N)^2$ by the maximum possible, we get:
\begin{align}
 \frac{(S/N)^2[W(ks_r)]}{(S/N)^2[W(s_r)]}&= \frac{\frac{T_0^2}{\pi\sigma^2s_r^2} \frac{s_{u}^{2}s_r^2}{(s_{u}^{2}+s_{r}^{2} )^{2}}}{\frac{T_0^2}{\pi\sigma^2s_r^2} \frac{s_{r}^{4}}{(s_{r}^{2}+s_{r}^{2} )^{2}}} \\&= 4\frac{s_u^2s_r^2}{(s_u^2 + s_r^2)^2} = \frac{(2k)^2}{(k^2+1)^2} = \left(\frac{2k}{k^2+1}\right)^2\,.
\end{align}
Figure~\ref{fig:SurveySpeed} shows a plot of the loss in survey speed when using the wrong PSF, as derived above.
This simple exercise gives us an idea how much sensitivity we lose when we combine images with different PSFs, because when we match filter the coadded image, all images are match filtered with a single PSF, that cannot simultaneously be optimal to all of them.

\end{document}